\documentclass[useAMS,usenatbib,twocolumn]{mn2e}

\usepackage[dvipdfmx]{graphicx}
\usepackage{amsmath,amssymb,natbib,siunitx,color}

\usepackage{color}

\input epsf
\usepackage{graphicx}
\newcommand{\beq}{\begin{equation}}
\newcommand{\beqa}{\begin{eqnarray}}
		  \newcommand{\eeq}{\end{equation}}
\newcommand{\eeqa}{\end{eqnarray}}

\newcommand{\lsim}{\lesssim}
\newcommand{\gsim}{\gtrsim}

\newcommand{\lmk}{\left(}
\newcommand{\rmk}{\right)}
\newcommand{\lnk}{\left\{ }
\newcommand{\rnk}{\right\} }
\newcommand{\lkk}{\left[} 
\newcommand{\rkk}{\right]}
\newcommand{\lla}{\left\langle}

\newcommand{\rra}{\right\rangle}

\newcommand{\mch}{{\cal M}}

\title{Orbital Synchronization Capture of Two Binaries Emitting Gravitational Waves
}

\author[N. Seto]{Naoki Seto
\\
Department of Physics, Kyoto University, 
Kyoto 606-8502, Japan\\
Center for Gravitational Physics, Yukawa Institute for Theoretical
Physics, Kyoto University, Kyoto 606-8502, Japan
}

\date{\today}

\begin{document}
\maketitle
\begin{abstract}
We study the possibility of orbital synchronization capture for a hierarchical quadrupole stellar system composed by
 two binaries emitting gravitational waves. Based on a simple model including the mass transfer for white dwarf binaries, we
 find that the capture might be realized for inter-binary distances less than their gravitational wavelength. 
 We also discuss  related intriguing phenomena such as  a parasitic relation between the coupled white dwarf binaries and significant reductions of gravitational and electromagnetic radiations. 

\end{abstract}


\section{introduction}

Synchronizations of multiple oscillators are  widely observed in physics and also in other fields such as biology, chemistry, engineering  and even social science \citep{pikovsky2003}. Historically, Huygens is considered to be  the first person who discovered and studied a synchronization phenomenon, using two pendulum clocks coupled through a wooden beam.  In 1665, he reported that the two clocks   settled into a synchronization state and emitted sounds simultaneously. In the 19th century, Rayleigh found that, depending on their mutual configuration,  two organ-pipes could almost reduce one another to silence, because of the mutual suppression of the oscillations \citep{rayleigh,abel2009}. Interestingly, in these classical examples, synchronization could be clearly identified by hearing sound waves.

Meanwhile, in 2015, the two advanced-LIGO detectors succeeded to catch gravitational waves (GWs) from merging black hole binaries \citep{ligovirgo2016}. GWs are emitted by moving celestial bodies and binaries are the most promising generators.  By carefully listening to GWs, we can inversely probe the dynamics of the GW sources and their  basic properties. 
The ground-based detectors such as the advanced LIGO have sensitivity to GWs  above $\sim 10$Hz, but, in the near future, a new window will be opened around 0.1mHz-0.1Hz by the Laser Interferometer Space Antenna (LISA,  \citealt{amaro-seoane2012}).

In this paper, considering  the ubiquitous emergence of synchronization,  we discuss whether two GW emitters come to have the same frequency (synchronization capture).   Paying attention to the impacts of mass transfer within white dwarf (WD) binaries, we specifically  discuss  synchronization of two binaries that are gravitationally coupled,  rotating each other as a quadrupole stellar system. 

For clarifying the fundamental mechanism of a synchronization process, it has been generally advantageous to simplify the model as much as possible, keeping only the essential degree of freedom \citep{pikovsky2003}. 
In this spirit, we employ the very basic model of \cite{p1972} for the mass transfer within each WD binary, and derive  ordinary differential equations to describe the time evolution of the gravitationally coupled binaries both with and without the mass transfer.  
We found that the mass transfer might provoke the synchronization capture and could resultantly cause intriguing phenomena such as a parasitic relation between the two WD binaries and a significant reduction of gravitational radiation due to a phase cancellation.

Note that, even if isolated, our binaries slowly change their rotation periods due to the gravitational radiation reaction. This is somewhat different from typical \lq\lq{}self-sustained oscillators\rq\rq{} that have intrinsic periods  like a pendulum clock \citep{pikovsky2003}.

Our Milky Way galaxy has a large number of mass-transferring WD binaries. For example,  $\sim 10^3$ of them might be found with electromagnetic (EM) telescopes ({\it e.g.} in the optical and X-ray bands) and $\sim 10^4$ might be detected with LISA \citep{nelemans2004}.  But, unfortunately,  it would be difficult to solidly estimate how many of them actually have nearby companion binaries relevant for synchronization capture, even though $\sim 10\%$ of solar-type binaries are expected   to have more than two companions \citep{raghavan2010}.  This is  partly because
the  binary evolution  in the earlier phases ({\it e.g.} during a giant star) is quite complicated.
In addition, as we see later, there could be a strong selection bias for observation of a synchronized  quadrupole system.  
Therefore, below, we concentrate on the dynamical aspects of the system.

This paper is organized as follows. In \S2,  we present the basic equations for evolution of two coupled binaries. 
In \S3, we study the synchronization capture for two inspiraling binaries without mass transfer such as two binary black holes. The capture is failed for this case. Then, in \S4, we demonstrate that the mass transfer within WD binaries can assist the capture, and explain the underlying mechanism.  
In \S5, we discuss the phase cancellation of GWs emitted from two synchronized WD binaries, and mention related observational implications. In \S6 we summarize our study and make a brief discussion.

\section{angular momenta of two binaries}

As shown in Fig.1, we study a quadrupole stellar system composed by two inner binaries (primary and secondary) that have semimajor axes $a_p$ and $a_s$, and are hierarchically  separated at the distance $d\ (\gg a_p, a_s)$.  For simplicity of calculation, the three orbits are assumed to be coplanar   and circular (partly because of the efficient energy dissipation processes especially  for the inner mass-transferring binaries). But we expect that qualitatively similar results will be obtained for inclined configurations, at least for small misalignment angles.   

In relation to our geometrical assumptions,  we mention the triple system recently identified through the timing analysis of the millisecond pulsar  PSR J0337+1715  \citep{ransom2014}. This system is composed by an inner neutron star-WD binary and an outer WD with nearly coplanar orbits (only $\sim0.2^\circ$ mismatch).  This alignment is considered to be caused by  dissipative processes during the formation of the outer WD.  A similar mechanism might be responsible for the configuration of our coplanar quadrupole system. 

Below, in \S 2.1, we discuss the orbital angular momentum and its evolution for isolated binaries, also including the effects of the mass transfer within binaries. Then, \S 2.2, we study the relevant torques induced by the binary-binary interaction.

\begin{figure}
 \includegraphics[width=7.5cm,angle=270]{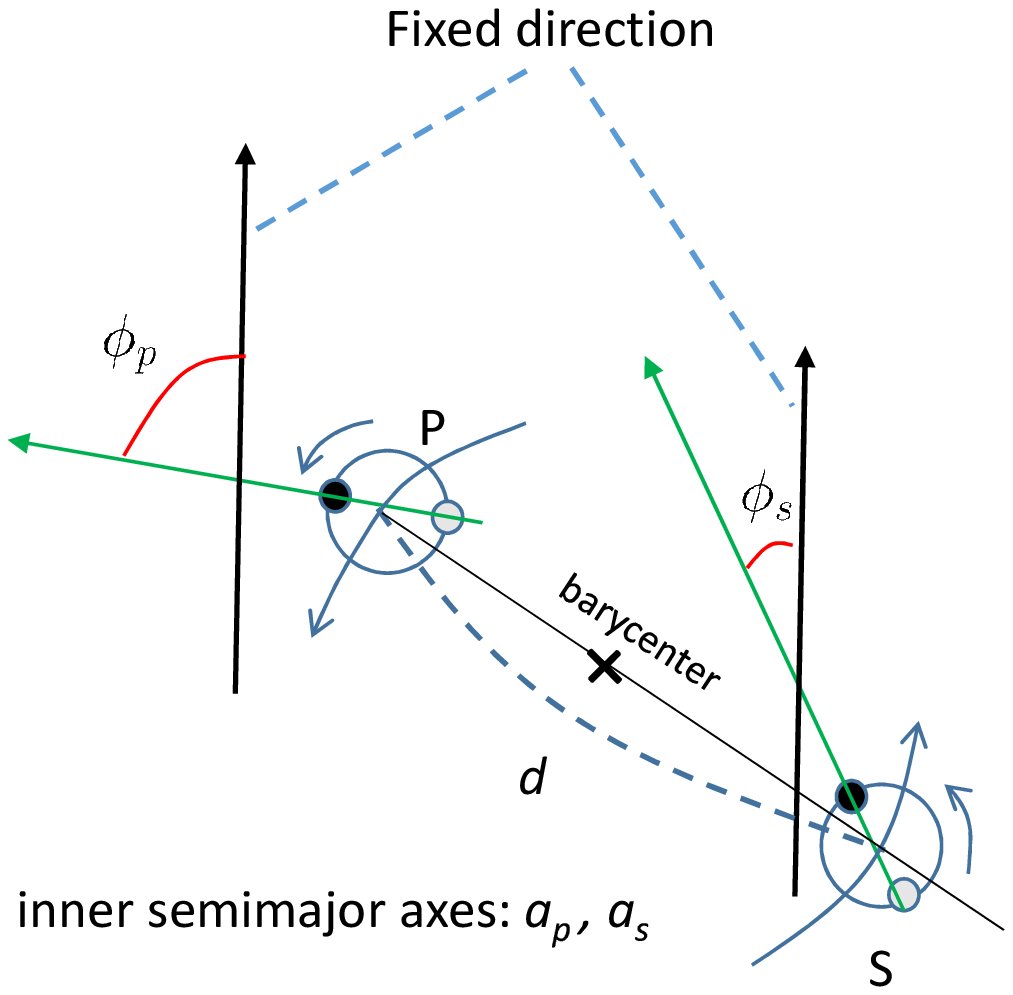}
 \caption{The geometry of two binaries on the same orbital plane. For the primary binary, we put the orientation angle $\phi_p$  relative to the fixed direction.  The angle $\phi_s$ for the secondary is defined in the same manner.  We denote the relative angle $\Delta\equiv \phi_s-\phi_p$. The inner semimajor axes are $a_p$ and $a_s$, and the outer orbital distance  is $d$. }
 \label{fig:geo}
\end{figure}

\subsection{Orbital angular momentum of each binary}

\subsubsection{original definition}

For the primary inner binary, we put their masses $m_{p1}$ and $m_{p2}$, and denote its total mass 
$M_p\equiv m_{p1}+m_{p2}$, reduced mass $\mu_p\equiv m_{p1}m_{p2}/M_p$ and  mass ratio $q_p\equiv m_{2p}/m_{1p}\le 1$. 
As shown in Fig.1,  we define the orientation angle $\phi_p$ of the binary (the vector from $m_{p1}$ to $m_{p2}$) relative to a fixed direction on the common orbital plane. 
 The orbital angular velocity is given by
$
n_p\equiv \dot{\phi}_p=\lmk{{GM_p}/{a_p^3}}\rmk^{1/2}$, and the frequency and the wavelength of the associated GW are given by 
$f_p=n_p/\pi$ and $\lambda_p=c \pi/n_p$. The orbital angular momentum of the primary binary is written as 
\beq
J_p=\mu_p (G M_p a_p)^{1/2}=\mu_p a_p^2 n_p.
\eeq
Here the product $\mu_p a_p^2$ is the quadrupole moment of the primary binary and would be frequently used in this paper. We have the following convenient relation  
\beq
\mu_p a_p^2=G^{2/3}\mch_p^{5/3}n_p^{-4/3}
\eeq
with the chirp mass
$\mch_p\equiv \mu_p^{3/5} M_p^{2/5}$.  We introduce the similar notations to the secondary binaries with the subscript $s$. The two binaries are distinguished by the condition  $\mch_{s}<\mch_p$.

Below, following  \cite{p1967}, we only deal with conservative mass transfer within each binary ({\it i.e.} $\dot{M}_p=\dot{M}_s=0$), including the cases without mass transfer ($\dot{m}_{p2}=\dot{m}_{s2}=0$) {\it e.g.} for binary black holes. Then we have 
\beqa
\frac{{\dot J}_p}{J_p}&=&\frac12 \frac{{\dot a}_{p}}{a_p}+\frac{{\dot m}_{p2}}{m_{p2}} (1-q_p),\\
\frac{{\dot J}_s}{J_s}&=&\frac12 \frac{{\dot a}_{s}}{a_s}+\frac{{\dot m}_{s2}}{m_{s2}} (1-q_s).\label{dj}
\eeqa

Due to the gravitational radiation reaction at the 2.5 post-Newtonian (PN) order, if isolated, the secondary  binary loses angular momentum and receives a negative torques given by (see {\it e.g.} \citealt{maggiore2008})
\beq
({\dot J}_s)_{gw}=Y_{ss}=-\frac{32 Ga_s^4 n_s^5 \mu_s^2}{5c^5}. \label{jgws}
\eeq
The characteristic evolution timescale is given by
\beqa
{t_{gw,s}}&\equiv& -\lkk \frac{({\dot J}_s)_{gw}}{J_s} \rkk^{-1}=\frac{5c^5}{32 G^{5/3} \mch_s^{5/3} n_s^{8/3}} \label{tgw}\\
&=&3.4\times 10^8 \lmk  \frac{\mch_s}{1M_\odot}\rmk^{-5/3}\lmk\frac{n}{\rm 0.001s^{-1}}\rmk^{-8/3}{\rm yr}.\nonumber
\eeqa
Similarly, for the primary binary parameters, we have
\beq
({\dot J}_p)_{gw}=Y_{pp}=-\frac{32 Ga_p^4 n_p^5 \mu_p^2}{5c^5}.\label{jgwp}
\eeq 
We should stress that the expressions (\ref{jgws})-(\ref{jgwp}) are given for isolated binaries. 
When two binaries are gravitationally coupled, we have additional torques, as discussed on \S 2.2.

\subsubsection{effects of mass transfer within each WD binary}
In \S 4, for the synchronization capture, we study the impacts of mass transfer within  double WD binaries. 
We use the very basic model of mass transfer developed by Paczy\'{n}ski \citep{p1967,p1972} (see also \citealt{gokhale2007} for a recent study) to follow the long-term evolution of orbital angular momentum. 
For notational simplicity, we temporally drop the subscripts $p$ and $s$.

We assume that the donor WD has a mass $m_2\ (<m_1)$ and satisfies the mass-radius relation for the polytropic   index 3/2
\beq
R_2=0.0126 R_\odot \lmk \frac{m_2}{1M_\odot}\rmk^{-1/3}.\label{mr}
\eeq
  For the Roche lobe radius, we use
\beq
R_{L2}=\frac{2a}{3^{4/3}} \lmk \frac{m_2}{m_1+m_2}   \rmk^{1/3}.\label{rl}
\eeq
Then the mass transfer from $m_2$ to $m_1$ is stable for $q\equiv m_2/m_1<2/3$, and we employ the rate 
\beq
\frac{{\dot m}_{2}}{m_{2}}=-2n \lmk \frac{R_{2}-R_{L2}}{R_2}  \rmk^3\theta(R_{2}-R_{L2}) \label{dm}
\eeq
given in \cite{p1972} (the factor 2 taken from \citealt{webbink1984}) and determined by the size of the donor $R_2$ relative to its Roche lobe radius $R_{L2}$.  Here, $\theta(\cdot)$ is the step function and the mass transfer does not occur when the Roche lobe is not filled with $R_2<R_{L2}$.

A mass-transferring WD binary can be formed after  filling the donor's Roche lobe in the last stage of the  orbital inspiral due to gravitational radiation.  Then, its separation $a$ starts to increase (decreasing the angular velocity $n$), compensating the radiative angular momentum loss by the mass transfer from the donor $m_2$ to the accreter $m_1$ [see eqs. (\ref{dj}) and (\ref{jgws})].   Therefore, eq.(\ref{dm}) can be regarded as the fuel consumption rate to generate angular momentum.

As we see later in \S 4, a mass-transferring WD binary has a simple and robust mechanism towards the synchronization capture.  Roughly speaking, when the orbital angular momentum is externally added to a binary, it reduce the fuel consumption rate $|\dot{m}_2|$ and weaken the decrease of $n$.  A similar but inverse process works, if angular momentum is extracted from the binary.  These responses stabilize the mass transfer and also assist  the synchronization capture for two  gravitationally-coupled binaries.

 At the quasi-steady state with $\ddot{m}_2\simeq0$, we have \citep{gokhale2007}
\beq
\frac{\dot a}{a}=-\frac23\frac{{\dot m}_2}{m_2}=\lmk1-\frac32q  \rmk^{-1} t_{gw}^{-1}. \label{iso}
\eeq
Here $t_{gw}$ is  same as that defined in eq.(\ref{tgw}), while the subscript $s$ is dropped here.

\subsection{Interaction between two binaries}
Now we discuss gravitational interaction between the two inner binaries. In \S2.2.1, we discuss the direct Newtonian interaction between them as the leading order conservative interaction. Then, in \S2.2.2, we mention the torques induced by the coherence of gravitational emissions, in addition to the already introduced expressions (\ref{jgws}) and (\ref{jgwp}) for isolated binaries. These   torques related to GWs are the leading (2.5PN) order dissipative terms, extracting the angular momentum from the system.  We make related discussions in \S2.2.3 and 2.2.4.

\subsubsection{torques by the mutual Newtonian interaction}

We are interested in the evolution  of two binaries around a synchronization capture where the angle $\Delta\equiv \phi_s-\phi_p$ changes slowly with time (${\dot \Delta}=\dot{\phi_s}-\dot{\phi_p}=n_s-n_p\simeq0$ because of the definition of the synchronization).
In our model, for the tidal torque from the primary to the secondary,  we simply apply the coherent term  given by \footnote{This expression is a factor of 9/21 different from \cite{batygin2015}. We can  evaluate the torque by directly considering interaction between four point masses (as shown in Fig. 1) and then perturbatively expanding $a_s/d$ and $a_p/d$ up to $(a/d)^4$, using a software such as {\it Mathematica}.  }
\beqa
T_s&=&-\frac{9G a_p^2 a_s^2 \mu_p \mu_s}{16d^5} \sin (2\Delta),
\eeqa
dropping the incoherent terms including that corresponding to the spin-orbit coupling   ({\it i.e.} the  interaction between  the inner and outer orbits, see also \citealt{batygin2015}). Here we assumed $d\ll\lambda_p\sim\lambda_s$, and ignored the time retardation associated with  the propagation of gravitational interaction. In the terminology of post-Newtonian analysis, the whole system is within the near zone \citep{maggiore2008}. 
The reaction torque of the primary is  $T_p=-T_s$.

\subsubsection{torques by the coherent GW emission}
As mentioned in \S 2.1.1, the two binaries receive negative torques  (\ref{jgws}) and (\ref{jgwp}) due to gravitational radiation reaction, if they are isolated.  However, around the orbital synchronization, we need to properly take into account  the coherence of the radiation for $d\ll \lambda_p\sim\lambda_s$ (see {\it e.g.} \citealt{maggiore2008}). We can roughly understand the reason in the following manner. At the lowest order of the PN expansion, the strain of gravitational wave $h$ is given by the second time derivative of the quadrupole moment of the system. But the angular momentum (and also energy) flux is proportional to $h^2$ (except for derivatives). Therefore, when the two binaries are nearby ($d\ll \lambda_p\sim \lambda_p$) and have similar orbital frequencies, the cross term of their quadrupole moments  could have a non-negligible contribution, after taking a time average.  
This coherent effect correspondingly  appears as radiative reaction forces and thus  associated torques.

 In order to include the coherence of gravitational radiation reaction, we employ the Burke-Thorne potential that is eventually given by the fifth time derivative of the total quadrupole moment (\citealt{thorne1969,burke1971} see also \citealt{maggiore2008} in the framework of the post-Newtonian analysis).  
In addition to the torque (\ref{jgws}) given for an isolated binary,  the secondary binary  has the new component $Y_{sp}$ induced by the primary
\beq
Y_{sp}=-\frac{32 Ga_s^2 a_p^2 n_p^5 \mu_p \mu_s}{5c^5} \cos(2\Delta).
\eeq
Similarly, the primary has the torque $Y_{ps}$ by the secondary
\beq
Y_{ps}=-\frac{32 Ga_p^2 a_s^2 n_s^5 \mu_p \mu_s}{5c^5} \cos(2\Delta).
\eeq
As mentioned earlier,  these expressions are valid for $d\ll \lambda_p\sim \lambda_s$.
For  $d\gg \lambda_p\sim \lambda_s$, the coherent effect of the GW emission becomes negligible, after  taking the surface integral of the angular momentum flux.

\subsubsection{final expressions for the torques}
Hereafter, whenever possible, for notational simplicity, we put $n_p=n_s=n$ (also $\lambda_p=\lambda_s=\lambda$), but,  if necessary,  appropriately handle the difference between $n_p$ and $n_s$.  Then, for the secondary binary,  we have
\beqa
\frac{{\dot J}_s}{J_s}&=&\frac{Y_{ss}+Y_{sp}+T_s}{J_s}\\
&=&-\frac1{t_{gw,s}} [1+F\cos(2\Delta)+DF\sin(2\Delta)]\label{djs}.
\eeqa
Here, we put
$F\equiv (\mch_p/\mch_s)^{5/3}\ge 1$
and 
$t_{gw,s}$ is  defined in eq. (\ref{tgw}) for a (hypothetically) isolated secondary binary. The factor $D$ is given by 
\beqa
D&\equiv& \frac{45 c^5 n^{-5}}{512d^5}=\frac{45 }{512\pi^5} \lmk\frac{\lambda}{d}\rmk^5\\
&=&36.2 \lmk \frac{n}{\rm 0.002 s^{-1}} \rmk^{-5} \lmk \frac{d}{\rm 0.3AU}  \rmk^{-5}\nonumber
\eeqa
and characterizes the strength of the direct gravitational coupling $T_s$ relative to the radiative one $Y_{ss}+Y_{sp}$.  The prefactor $45/512/\pi^5\sim1/3500$ is much smaller  than unity but we are assuming $\lambda\gg d$.  
Below, we mainly consider the ranges  $(F-1)\sim 0.1$ and $5\lsim D\lsim 50$. The latter reflects the validity of our expressions ($\lambda\gg d$) and reality of the system (though somewhat arbitrary). Note also that the factor $(F-1)$ is unlikely to be very small ({\it e.g.} 0.01), considering the occurrence of the catch up $\dot \Delta=0$ for two evolving binaries.

 In the same manner, for the primary binary,  we obtain
\beqa
\frac{{\dot J}_p}{J_p}&=&\frac{Y_{pp}+Y_{ps}+T_p}{J_p}\\
&=&-\frac{1}{t_{gw,s}} [F+\cos(2\Delta)-D\sin(2\Delta)].\label{djp}
\eeqa

\subsubsection{other effects}

In terms of the post-Newtonian analysis, we have included the Newtonian order for the conservative terms and the 2.5PN order for the dissipative ones, both as the leading order effects.  Since the angular velocities $n_p=\dot{\phi_p}$ and  $n_s=\dot{\phi_s}$ is mainly determined by the Newtonian dynamics, the conservative 1PN and 2PN effects would be insignificant for the evolution of $\Delta=\phi_s-\phi_p$. But these conservative higher PN effects sometimes become important in celestial mechanics, as we briefly mention below.

 In relation to our study, one might interested in the Kozai mechanism that could oscillates the inner eccentricity and inclination of   a hierarchical system with mutually inclined orbits (see {\it e.g.} \citealt{hamers2017,fang2017}).   In contrast to the purely Newtonian dynamics in which  a binary (of two point particles) moves on a fixed elliptical orbit, the 1PN effect precesses the pericenter  with  the timescale $t_{1PN}\sim n_s^{-1} c^2a_s/(G M_s)$ ({\it e.g.} for the secondary).  Meanwhile, the precession timescale of the    Kozai effect is $t_K\sim n_s^{-1}(d/a_s)^3$ \citep{holman1997}.  Then we  have
\beq
\frac{t_{K}}{t_{1PN}}\sim \frac1{3D^{2/5}}\frac{d}{a_p}
\eeq
that is much larger than unity for the masses and orbital separations of the point particle system analyzed in the next section. 
Therefore, although we basically consider the coplanar systems, the Kozai mechanism is suppressed by the 1PN effect even for inclined cases \citep{holman1997}, at least around the synchronization.  But further study might be required {\it e.g.} for the earlier stages.

\section{inspiral binaries without mass transfer}

We first examine the interaction between two inspiral binaries effectively made by point particles such as black holes and neutron stars without mass transfer (${\dot m}_{p2}={\dot m}_{s2}=0$).  Actually,  these relatively simple systems will turn out to be unpromising for realizing orbital synchronization, but would be quite useful to elucidate the key physical processes.

As a concrete model, we consider the following quadrupole system composed by two black hole binaries of masses $10 M_\odot+8M_\odot$ and $10 M_\odot+6.95M_\odot$  ($F=1.075$).  At $t=0$, we set $\Delta=0$, $n_s=0.002{\rm s^{-1}}$ ($a_s=8.28\times 10^{10}$cm) and $n_p=0.9999n_s$ ($a_p=8.44\times 10^{10}$cm) with their separation $d=4.5\times 10^{12}{\rm cm}=0.3{\rm AU}$.  For these parameters we have the timescale  $t_{gw,s}=2.0\times 10^6$yr and the coupling parameter $D=36.2$.  The orbital parameters $(a_p,a_s,d)$ well satisfy the stability condition for hierarchical orbits \citep{mardling2001}.

As we initially have $n_s>n_p$ ({\it i.e.} ${\dot \Delta}>0$), the primary catches up the secondary to have  ${\dot \Delta}=0$ at the turnover epoch $t=t_T=628$yr with the angle $\Delta =\Delta_T=2805.5$ (see Fig. 2). Since then, the angle $\Delta$ continues to decrease. The  synchronization capture ($\dot{\Delta}\sim 0$ {\it i.e.} $\Delta\sim const$) was not realized in the present model.
Actually this failure is not specific to our model parameters, but could be widely expected for the simple quadrupole systems made by point particles, as we see analytically below.

For the synchronization capture, the time evolution of  $\Delta$  should be carefully examined during the two consecutive phases $[\Delta_T-\pi,\Delta_T]$ and $[\Delta_T, \Delta_T-\pi]$ around the turnover epoch $t=t_T$. We showed this critical turnover period in Fig.2.   
From eqs. (\ref{dj})(\ref{djs})(\ref{djp}) as well as the relation 
\beq
{\ddot \Delta}={\dot n}_s-\dot{n}_p=-\frac32n \lmk \frac{\dot a_s}{ a_s}- \frac{\dot a_p}{ a_p} \rmk, \label{b1}
\eeq
 we have
\beq
{\ddot {\Delta}}=\frac{3n}{t_{gw,s}} \lkk (1-F) (1-\cos2\Delta)+D(F+1)\sin2\Delta\rkk . \label{b2}
\eeq
Note that, in the third expression in eq. (\ref{b1}), we dropped the term that is $O[(n_s-n_p)/n]$ times smaller than the right-hand side of eq. (\ref{b2}).
Here, in our model,  the parameters $(n,t_{gw,s},D)$ change slowly with a timescale at least $t_{gw,s}/D$ (including the effects for the simplification $n_p=n_s=n$). This is much longer than the characteristic duration    $t_{lib}=(6 t_{gw,s}/nD)^{1/2}$ of the critical turnover period.  Therefore, temporarily ignoring their time dependences, we get  an useful approximate relation
\beq
\frac12 {\dot \Delta}^2+ \frac{3n}{2t_{gw,s}} \big[ (F-1) (2\Delta+\sin2\Delta)
 +D(F+1)\cos2\Delta\big]=E, \label{en}
\eeq
where the integral constant $E$ can be regarded as energy and the term proportional to the square bracket corresponds to the potential, dominated by $\propto \cos2\Delta$ (originating from the direct Newtonian torque) in the present setting with $(F-1)\ll 1$ and $D\gg 1$.  
Note that these terminologies (energy and potential) are expedientially introduced to analogically understand the structure of our effective equation (\ref{en}).  They are not simply related to the actual energy or potential of the original quadrupole system.

\begin{figure}
 \includegraphics[width=.9\linewidth]{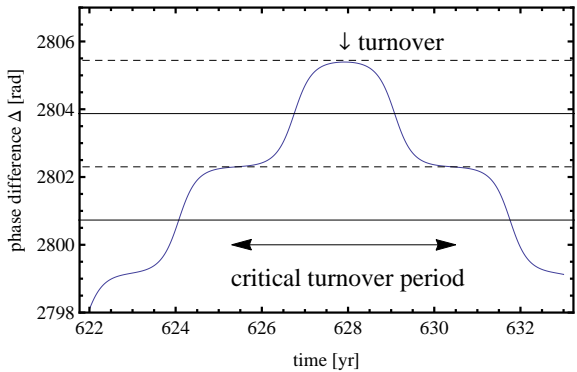}
 \caption{The evolution of the phase difference $\Delta$ around the turnover  at $t=t_T\sim 628$yr. This system is composed by two inspiraling black hole binaries (see \S 3).   The solid horizontal line represents the angle  with $\cos(2\Delta)=-1$, and the dashed ones are for $\cos(2\Delta)=1$.  They correspond to local minimums and maximums of the potential in eq. (\ref{en}). The synchronization captured was failed in this figure.  }
 \label{fig:distance}
\end{figure}

The potential has stable points (local minimums) around $\cos2\Delta=-1+2(1-F)^2/(F+1)^2/D^2$ [hereafter ignoring the correction $O(D^{-2})$],   but the existence of a stable point is just a necessary condition for the synchronization capture.   
During the critical turnover period, by some time irreversible processes ({\it e.g.} similar to frictional dissipation),  we additionally need to reduce the energy $E$ in the approximate relation (\ref{en}). 
The situation is analog to the traditional  spin-orbit resonant capture for a celestial body with a permanent quadrupole moment such as a satellite \citep{batygin2015,goldreich1968,ssd}. For a capture probability of order unity, the energy $E$ should be deceased by $\delta E\sim O[(F-1)n/t_{gw,s}]$ during the critical period.  



Even if we take into account the variation of the parameters $(n, t_{gw,s},D)$ by GW emission, the energy variation (apart from its sign)  in eq.(\ref{en}) is estimated to be at most $O[n^{1/2}D^{3/2}t_{gw,s}^{-3/2}]$ that is generally much smaller than the required level $\delta E$. 
Therefore, the gravitational radiation alone would not be sufficient to realize the synchronization capture. 

If the two binaries involve WDs, their spin might assist the capture. But, in the next section, we show that the effects of mass transfer could provoke the capture.

\section{outspiral WD binaries with mass transfer}

We now discuss the time evolution of two mass-transferring WD binaries, gravitationally coupled at the distance $d$. As in the previous section, using eqs.(\ref{dj})(\ref{djs})(\ref{djp}) and (\ref{dm}), we obtain ordinary differential equations for the five variables $\Delta$, $a_p$, $a_s$, $m_{p2}$ and $m_{s2}$. 
More specifically, we have 
\beq
\frac12 \frac{{\dot a}_{p}}{a_p}+\frac{{\dot m}_{p2}}{m_{p2}} (1-q_p)=-\frac{1}{t_{gw,s}} [F+\cos(2\Delta)-D\sin(2\Delta)],\label{djp2}
\eeq
\beq
\frac12 \frac{{\dot a}_{s}}{a_s}+\frac{{\dot m}_{s2}}{m_{s2}} (1-q_s)=-\frac{1}{t_{gw,s}} [1+F\cos(2\Delta)+DF\sin(2\Delta)],\label{djp3}
\eeq
\beq
{\dot \Delta}=\lmk \frac{GM_s}{a_s^3}  \rmk^{1/2}-\lmk \frac{GM_p}{a_p^3}  \rmk^{1/2},
\eeq
and two equations corresponding to eq.(\ref{dm}) for the mass transfer rates both for the primary and the secondary binaries. 
{Here, we again ignore the effects of short-period terms for which further study might be required.}

\begin{figure}
 \includegraphics[width=.95\linewidth]{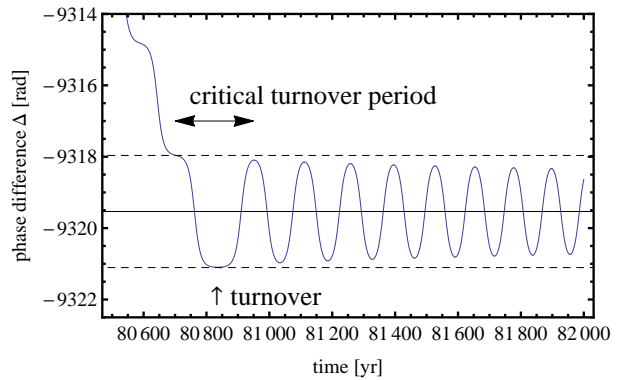}
 \caption{The evolution of the phase difference $\Delta$ around the synchronization capture  at $t\sim 8.1\times 10^4$yr. The system is composed by two mass-transferring WD binaries (see \S 4).  The solid horizontal line represents the angle  $\Delta =-9319.53$ with $\cos(2\Delta)=-1$, and the dashed ones are for $\cos(2\Delta)=1$.  They correspond to the local minimums (stable points) and the local maximums of the potential in eq. (\ref{en}).  The energy dissipation during the critical turnover period ($\sim 80700$yr to $\sim 80950$yr) around the first turnover at $t_T=80840$yr is essential for the capture. }
 \label{fig:distance}
\end{figure}

To be concrete, we performed  numerical integration of these differential equations  from the following fiducial initial conditions at $t=0$; $\Delta=0$,  $M_p=1.0M_\odot$, $M_s=0.9M_\odot$, $n_s=0.002{\rm s^{-1}}$,  and $d=0.3$AU (corresponding to $D=36.2$,  and  $t_{gw,s}=4.1\times 10^9$yr). Note that the donor masses\footnote{For this small mass, the mass-radius relation could have an index ({\it e.g.} $-0.2$) smaller than $-1/3$ used in our simplified model eq. (\ref{mr}) ({see} \citealt{verbunt1988}). }  $m_{p2}\simeq m_{s2}\simeq 0.0144M_\odot$  are roughly determined from eqs.(\ref{mr}) and  (\ref{rl}), given the angular velocity $n$.  More specifically, from $R_2\simeq R_{L2}$, we have $m_{p2}\simeq m_{s2}\propto n$ in our simple model.
We further tune the masses so that the initial mass-transfer rates ${\dot m}_{p2}$ and ${\dot m}_{s2}$ agree with eq. (\ref{iso}) originally given  for the isolated binaries. 
The initial mismatch of the angular velocities  is set at $n_p/n_s-1=3.0\times10^{-6}$ to ensure a large rotation cycles of $\Delta$ before the turnover, reducing the transients effects caused by our artificial initial settings.  The chirp masses here ($\mch_p=0.079M_\odot$ and $\mch_s=0.075M_\odot$ with $F=1.075$) are much smaller than the previous case in \S 3, and we have a much longer  evolution timescale $t_{gw,s}$. The orbital parameters $(a_p,a_s,d)$ again satisfy the stability condition in \cite{mardling2001}.

First, we discuss our numerical results. As shown in Fig. 3, at $t=t_T=80840$yr,  after $\sim 1500$ rotations of $\Delta$, the system reaches the turnover ${\dot \Delta}=0$ (with $\Delta_T=-9321.1$) for the first time, and was successfully captured into the orbit-orbit synchronization.  Since then,  its oscillation amplitude  gradually decreases down to the local minimum of the potential around $\Delta=-9318.5(\sim \Delta_T+\pi/2)$.

 Next, as in the previous case in \S 3, we analytically discuss  the capture process with the approximate relation  (\ref{en}).  Including the mass transfers, its right-hand-side is modified as 
\beq
E+3n\int_{t_T}^t dt \lnk \frac{\dot{m}_{s2}}{m_{s2}}(1-q_s)- \frac{\dot{m}_{p2}}{m_{p2}}(1-q_p) \rnk {\dot \Delta},\label{fri}
\eeq
and the energy of the approximate relation  (\ref{en}) should be now regarded as time dependent. 
During and after the critical period around the capture, mainly due to the modulations of $a_{p}$ and $a_s$ caused by the mutual torque $T_s=-T_p$, the rates ${\dot m}_{p2}$ and ${\dot m}_{s2}$ fluctuate around their mean values.  Here, the fluctuation $\delta {\dot m}_{p2}$ is in phase with $\dot \Delta$, but  $\delta {\dot m}_{s2}$ is in anti-phase. This frictionally reduces the oscillation energy (\ref{fri})  of $\Delta$, helps the  synchronization capture and settles the angle $\Delta$ down into the bottom of the potential. Indeed, {for $D\gsim 1$}, assuming the mean rates  $\lla {\dot m}_{p2}/m_{p2}\rra \sim \lla {\dot m}_{s2}/m_{s2}\rra\sim -t_{gw,s}^{-1}$ around the turnover, the dissipated energy during the critical turnover period is roughly estimated as $\sim 100n/t_{gw,s} \times D^{1/2} (n t_{gw,s})^{-1/6}$. For $D>1$, this is comfortably larger than the required level $\delta E=O[(F-1)n/t_{gw,s}]$ mentioned before.

Our coupling parameter so far was $D=36.2$ with the separation $d=0.3$AU.  We also examined our differential equations for larger $d$      (smaller $D$), keeping other parameters fixed but relaxing the requirement $d\ll \lambda$. For various $D$, we individually performed 10 numerical runs with different initial phases  $\Delta$.  We found that the capture rate was 100\% for $d\lsim 0.74$AU ($D\gsim 0.4$) but not $100\%$ for $d\gsim 0.74$AU.  In addition, the capture rate  was less than $\sim10\%$ for $d\gsim 1.1$AU.  These results are roughly consistent with what is expected from the energy balance argument around the first turnover \citep{ssd}. 
Namely, the condition for the  capture is given by  $100n/t_{gw,s} \times D^{1/2} (n t_{gw,s})^{-1/6}\gsim (F-1)n/t_{gw,s}$ or equivalently 
\beqa
d&\lsim& 4 (F-1)^{-2/5}c^{2/3}n^{-8/9} (G\mch_s)^{1/9} \nonumber\\
&\simeq& 1.2{\rm AU} \lmk \frac{F-1}{0.075}\rmk^{-2/5}\lmk \frac{m_{s2}}{\rm 0.0144 M_\odot}\rmk^{-37/45}\nonumber\\
& & \times\lmk \frac{M_{s}}{\rm 0.9 M_\odot}\rmk^{-2/45} .\label{ineq}
\eeqa
Here we used the relations $m_{s2}\propto n$ and $m_{s2}\ll m_{s1}\simeq M_s$ for the mass transferring WD binaries (see eqs.(\ref{mr}) and (\ref{rl})).  In eq.(\ref{ineq}) we also plugged-in our model parameters for the above numerical experiments. Note that the left-hand-side depends  very weakly on the total mass $M_s$ of the secondary, compared with the donor mass $m_{s2}$ (or equivalently the angular velocity $n$) and the ratio $F$.
{Strictly speaking, eq. (\ref{ineq}) is derived under the assumption  $D\gsim 1$.}

\section{ GW phase cancellation and observational implications}
In this section, we discuss interesting phenomena observed after the synchronization capture. The numerical results presented in Figs.4 and 5 are obtained for the same system as analyzed in \S 4. For this system, from $t=0$ to $2.5 \times 10^5$yr, the total variations $\delta a_j/a_j$ and $\delta m_{j2}/m_{j2}$ ($j=p,s$) are less than $10^{-4}$.

In Fig.4, we present the fuel consumption rates.
 The leftmost values are the rates for two isolated binaries without the coupling [see eq.(\ref{iso})]. The rates $|{\dot m}_{p2}|$ and $|{\dot m}_{s2}|$ increase until the capture epoch at $t\sim8.1\times 10^4$yr, but then start to decrease significantly.  These decrements would result in darkening the EM signals powered  by the mass accretion.

Actually, the shrinkage of the fuel consumption  rates is caused by the fall-off of the GW luminosity.  In Fig. 5, we present the normalized  GW luminosity 
\beq
P\equiv   \frac{ 1+F^2+2F\cos[2\Delta]}{1+F^2}\label{pow}
\eeq 
in our model. 
We can simply understand this expression in the following manner. As mentioned earlier, the GW strain $h$ of a single binary is proportional to its quadrupole moment (5/3th power of its chirp mass).  Additionally considering the wave phases, the time profile of  the total GW strain from the two binaries is given as
\beq
h(t)\propto \mch_p^{5/3}\cos[2\phi_p(t)]+ \mch_s^{5/3}\cos[2\phi_s(t)].\label{str}
\eeq 
Here we assumed a  separation $d\ll \lambda$.
The energy flux is roughly proportional to 
\beqa
h(t)^2&\propto&   \cos^2[2\phi_s(t)] +F^2 \cos^2[2\phi_p(t)]\nonumber\\
& &+F \lnk  \cos[2\phi_p(t)+2\phi_p(t)]+\cos[2\Delta(t)]  \rnk . \nonumber
\eeqa
Using the time averages $\lla \cos^2[2\phi_s(t)] \rra =\lla \cos^2[2\phi_p(t)] \rra=1/2$,  $\lla \cos[2\phi_p(t)+2\phi_p(t)] \rra=0$ and $\lla \cos[2\Delta(t)]\rra=\cos[2\Delta(t)]$ (given the slow variation of $\Delta(t)$ around and after the synchronization), we recover eq.(\ref{pow}), after including the normalization factor.
Note that we have the time averages  $\lla \cos[2\Delta]\rra=0$ and $\lla P\rra=1$ for two incoherent binaries, as for the early phase in Fig. 5.  After the  capture, the angle $\Delta$ frictionally damps towards the bottom of the potential at $\cos 2\Delta=-1$, where the GW luminosity becomes minimum, due to the  phase cancellation ($\cos[2\phi_p(t)]\simeq -\cos[2\phi_s(t)]$ in eq.(\ref{str})), or equivalently, cancellation of the quadrupole moment. The reduction factor of the GW luminosity is $\sim(F-1)^2$.  The actual luminosity could be somewhat larger, because of the finiteness of the gravitational wavelength $\lambda$.

\begin{figure}
 \includegraphics[width=.95\linewidth]{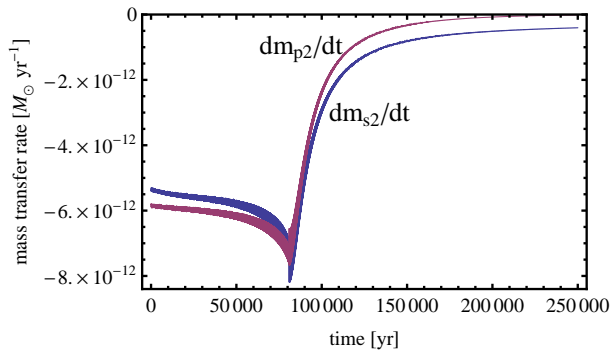}
 \caption{The fuel consumption (mass transfer) rates of two mass-transferring WD binaries.  After the synchronization at $t\sim8.1\times 10^4$yr, the magnitudes of rates decease significantly.  We  identically have ${\dot m}_{p2}=0$ for $t>3.4\times 10^5$yr.}
 \label{fig:distance}
\end{figure}

\begin{figure}
 \includegraphics[width=.95\linewidth]{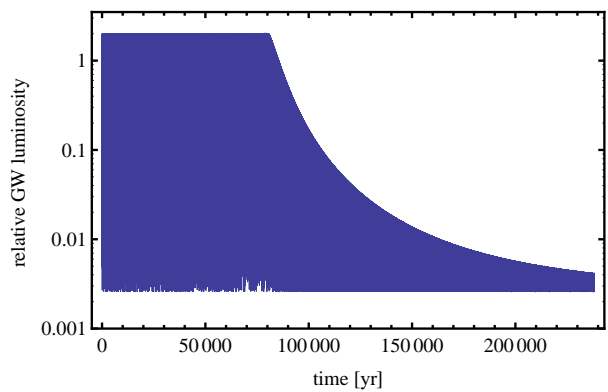}
 \caption{The luminosity of GW emission $P\equiv ( 1+F^2+2F\cos[2\Delta])/(1+F^2)$ from the mass-transferring WD binaries,  normalized by the luminosity of the incoherent binaries. For our model parameters $F=1.075$, the ratio $P$ takes its maximum   $\sim 2$ at $\cos 2\Delta=1$ and  minimum  $2.6\times 10^{-3}$ at $\cos 2\Delta=-1$. After the synchronization capture, the GW luminosity decreases, as the angle $\Delta$ settles down to the bottom of the potential at $\cos2\Delta=-1$. }
 \label{fig:distance}
\end{figure}

Interestingly,  in Fig. 4, the primary\rq{}s  fuel consumption rate ${\dot m}_{p2}$ approaches 0.  Indeed, we have ${\dot m}_{p2}=0$ for $t>3.4\times 10^5$yr.  Under the synchronization, by parasitizing  the secondary binary,  the primary increases its  angular momentum without consuming its own fuel. Depending on the initial conditions, the secondary can inversely pick up the angular momentum out of  the primary.  In fact, these uneven states are natural outcomes of the present mass-transfer process. If either binary happens to relatively decrease its fuel consumption rate under the synchronization, its difference $R_2-R_{2L}$ also becomes smaller  and could further reduce the rate, compared  with the counterpart.

Finally we discuss the asymptotic state after the oscillation of $\Delta$ becomes small. Using the conditions ${\dot m}_{p2}=0$, ${\dot \Delta}=0$ and ${\ddot m}_{s2}\simeq0$ \citep{gokhale2007},   we have
\beq
\frac{\dot{a}_s}{a_s}=\frac{\dot{a}_p}{a_p}=-\frac23\frac{\dot{m}_{s2}}{m_{s2}}. \label{asy}
\eeq 
From these relations, we can show that,
among the angular momentum lost from the secondary binary, the fraction 
\beq
F(2-3q_s)^{-1}\label{fra}
\eeq
 is stored in the primary binary and the rest is radiated away as GWs.

\section{summary and discussion}
Synchronization states are ubiquitously identified in various research  fields. When coupled, multiple oscillators can show intriguing  behaviors that are difficult to be anticipated from an isolated oscillator. In  this paper, using a simplified model, we discuss the evolution of two gravitationally coupled binaries that  emit GWs. We found that the mass transfer by the Roche lobe overflow of binary WDs could help the synchronization capture. This is due to the self-regulating mechanism of the WD binaries to keep the mass transfer stable. It effectively softens the response of orbital angular velocities to externally imposed torque, and resultantly assists the synchronization capture for two coupled binaries. Furthermore, it frictionally damps the relative angle $\Delta$ down to its stable point $\cos[2\Delta]\simeq-1$. 
From the energy balance argument \citep{ssd}, the separation $d$ between the two synchronized binaries should satisfy the inequality (\ref{ineq}). 

We also showed that, taking the advantage of the orbital synchronization,  one of the coupled WD binaries can start  absorbing the angular momentum of the counterpart.   Furthermore, the parasitic binary also tries to ably evade our search with EM and GW telescopes, by decreasing the mass transfer rate and the GW luminosity by a factor of $\sim(F-1)^2$. Correspondingly, the signal-to-noise ratio of the GW signal becomes $|F-1|$ times smaller than  the isolated binaries.  Here the coherence of the GW emission and its phase cancellation at the stable phase difference $\cos[2\Delta]\simeq-1$ play crucial roles.

 In this paper, we have considered the case $d\ll \lambda$ and neglected the time retardation for the binary-binary interaction. But, in reality, there could be corrections ({\it e.g.} for the relative phase $\Delta$) induced by the retardation of time varying gravitational field, and they would be interesting probes for the dynamical nature of gravitational interaction. Therefore, although the detectability of the synchronized WD binaries  is beyond the scope of this paper, they might become important observational targets in future.   In any case,  to separately determine various geometrical parameters, combination of  GW and EM observations would be particularly useful for studying these systems.   

So far, we have concentrated on physics around the synchronization capture. Once realized, the synchronization state  would be kept for a long period of time $\sim t_{gw,s}(F-1)^{-2}$ that could be even longer than the Hubble time. The subsequent evolution of the system would be also interesting. One possibility is that at some stage, the parasitized binary could not hold the synchronization state due to its reduced donor mass ({\it i.e.} increasing the ratio $F$).  Indeed, the fraction (\ref{fra}) becomes unphysical $\ge 1$ at $F\sim 2$.
In Fig.6,   we present a schematic diagram for the evolution of the two angular velocities in this  de-synchronization scenario, assuming that the primary binary  parasitizes the secondary at the earlier  epoch A.  
 In this figure, at the de-synchronization epoch B, the primary binary does not fill its Roche lobe and  would therefore start to inspiral again, increasing its angular velocity $n_p$.  Then, at the epoch C, when $n_p$ reaches the value at the synchronization capture A, the donor of the primary fills its Roche lobe, and resumes the mass transfer, shifting $n_p$ downward. Because of the relation $\mch_p>\mch_s$ after the epoch C,  the angular velocities can match ${\dot \Delta}=n_s-n_p=0$ at the epoch D.  Then the synchronization state might be recovered for the second time.

\begin{figure}
 \includegraphics[width=.70\linewidth,angle=270]{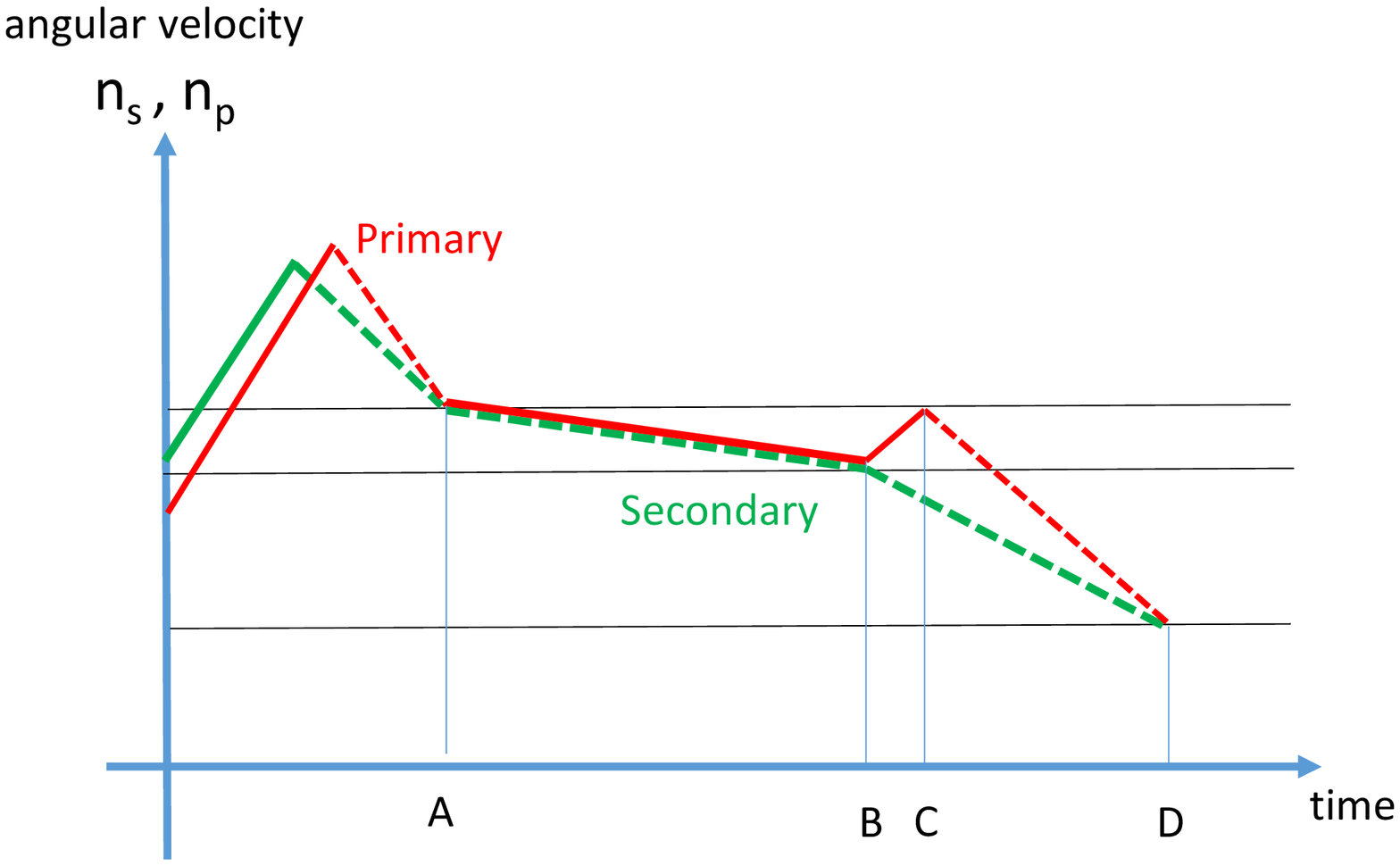}
 \caption{The schematic diagram for the evolution of the angular velocities  $(n_p,n_s)$ of the primary and secondary binaries. The dashed lines represent the phases with mass transfer and the solid lines are for those without mass transfer. The synchronization captures occur at the epoch A, and the primary is assumed to parasitize the secondary.     At the epoch B, the system escapes the synchronization state.  We have $n_p=n_s$ again at the epoch D.   }
 \label{fig:distance}
\end{figure}

\section*{Acknowledgements}

The author would like to thank the referee for helpful comments to improve the draft.
 This work is supported by JSPS Kakenhi Grant-in-Aid for Scientific Research
 (Nos.~15K65075, 17H06358).

\bibliographystyle{mn2e}


\end{document}